\documentclass{PoS}
\usepackage{siunitx}

\title{Neutrino direction and energy resolution of Askaryan detectors}

\ShortTitle{Neutrino direction and energy reconstruction}

\author{\speaker{Christian Glaser} for the ARIANNA collaboration\footnote{for collaboration list see PoS(ICRC2019)1177}\\
        University of California, Irvine, USA
        E-mail: \email{christian.glaser@uci.edu}}

\abstract{Detection of high-energy neutrinos via the radio technique allows for an exploration of the neutrino energy range from $\sim$\SI{e16}{eV} to $\sim$\SI{e20}{eV} with unprecedented precision. These \emph{Askaryan} detectors have matured in two pilot arrays (ARA and ARIANNA) and the construction of a large-scale detector is actively discussed in the community. In this contribution, we present reconstruction techniques to determine the neutrino direction and energy from the observed few-nanoseconds short radio flashes and quantify the resolution of one of such detectors. The reconstruction of the neutrino direction requires a precise measurement of both the signal direction as well as the signal polarization. The reconstruction of the neutrino energy requires, in addition, the measurement of the vertex distance, obtainable from the time difference of two signal paths through the ice, and the viewing angle of the in-ice shower via the frequency spectrum. We discuss the required algorithms and quantify the resolution using a detailed Monte Carlo simulation study.}

\FullConference{36th International Cosmic Ray Conference -ICRC2019-\\
		July 24th - August 1st, 2019\\
		Madison, WI, U.S.A.}

\begin{document}

\section{Introduction}

In this contribution, we study how the crucial neutrino properties direction and energy can be determined from a radio neutrino detector, a topic not yet studied thoroughly in the literature. The capabilities to reconstruct the neutrino properties are an important design criterion for future detectors. This paper discusses the experimental requirements and can be used to evaluate a detector design. 
If possible we keep our discussion general but if needed we will discuss a specific shallow station layout that has emerged out of the success of the ARIANNA pilot detector, is also used in the examples of NuRadioMC \cite{NuRadioMC}, and was already optimized for a good reconstruction performance \cite{COSPAR2018}.

\section{Sensitivity to neutrino direction and energy}
In this section, we first discuss how the measured radio signal relates to the neutrino properties and show which quantities of the Askaryan signal need to be measured to reconstruct the direction and energy of the neutrino. 

The dependence of the Askaryan signal arriving at the detector can be expressed in the following equation
\begin{equation}
    \vec{\varepsilon}(f) = \vec{e}_p(\vec{p}, \vec{v}_\nu, \vec{X}_\mathrm{int}) \times |\vec{\varepsilon}_0| \times \frac{e^{-R/L(f)}}{R} \times \exp\left[\frac{-(\theta - \theta_C)^2}{2 \sigma_\theta(f)^2}\right] \, .
    \label{eq:all}
\end{equation}
The four parts of the equation are discussed in the following.

\subsection{Signal direction and polarization}
\begin{figure}[t]
    \centering
    \includegraphics[width=0.5\textwidth]{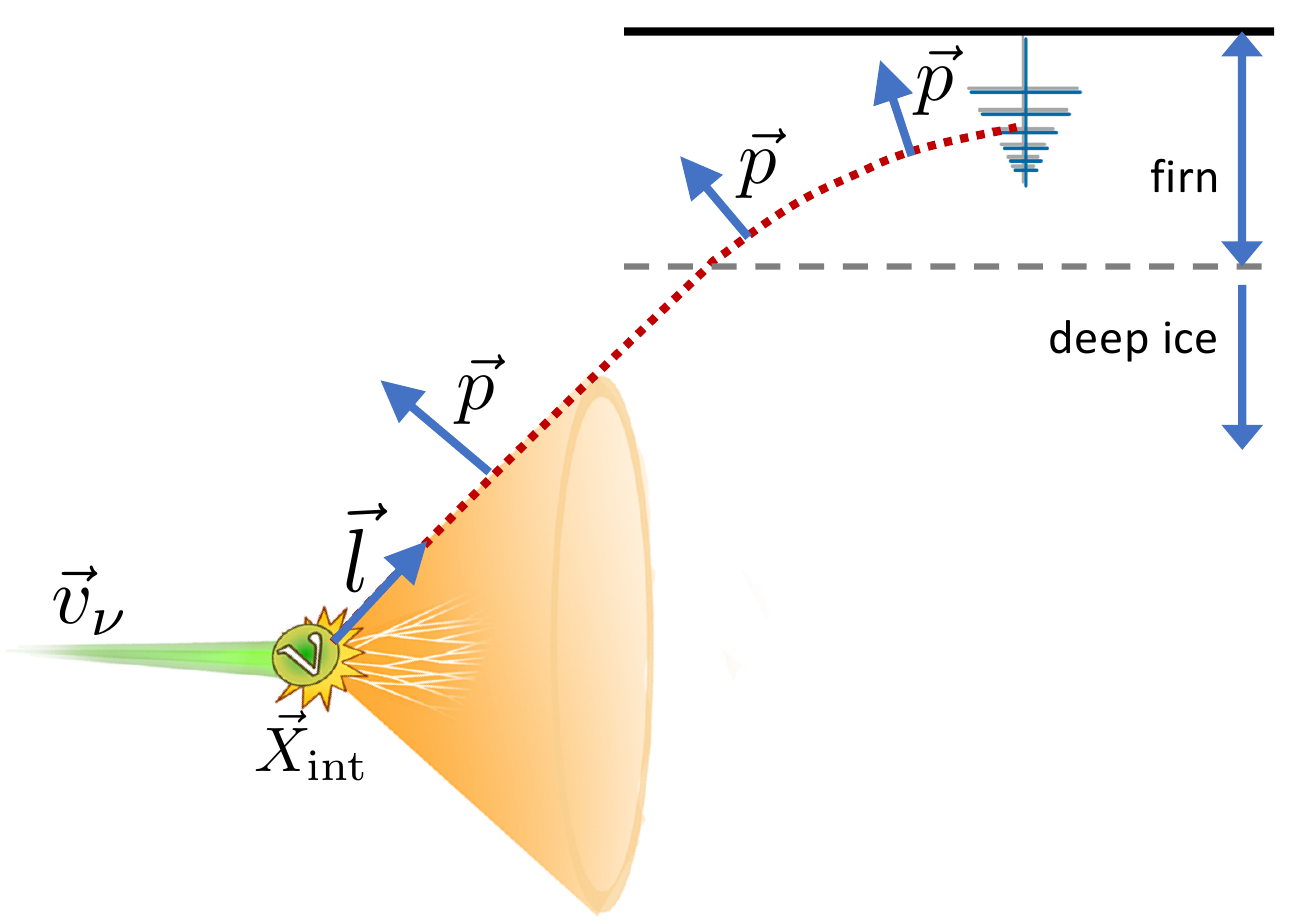}
    \caption{Polarization of the neutrino signal. Please note that the sketch is only a 2D projection of the problem.}
    \label{fig:polarization}
\end{figure}

The first part of Eq.~(\ref{eq:all}) describes the polarization of the Askaryan signal at the detector. The polarization of the Askaryan signal $\vec{p}$ at the point of emission is perpendicular to its direction of propagation (the launch vector $\vec{l}$) and the plane spanned by the neutrino direction $\vec{v}_\nu$ and the direction of signal propagation:
\begin{equation}
    \vec{p} = \vec{l} \times (\vec{v}_\nu \times \vec{l}) \, ,
    \label{eq:polarization}
\end{equation}
During propagation of the signal to the detector, the polarization is altered. In the upper layers of the ice -- the firn -- the index of refraction changes continuously from $n = 1.78$ of deep ice to $n \approx 1.35$ at the ice surface, which leads to continuous Fresnel refraction. The polarization changes according to the bending of the signal path to remain perpendicular to it. Thus, the polarization also depends on the position of the neutrino interaction vertex $\vec{X}_\mathrm{int}$ which determines the path between emitter and receiver. This dependence is illustrated in Fig.~\ref{fig:polarization}. 

To obtain the neutrino direction from a measurement of the polarization, Eq.~(\ref{eq:polarization}) can be solved for the neutrino direction 
\begin{equation}
    \vec{\hat{v}}_\nu = \sin \theta \vec{\hat{p}} - \cos \theta \, \vec{\hat{l}} \, ,
\end{equation}
where the $\hat{}$ symbol indicates that the vectors have unit length, and $\theta$ is the viewing angle, i.e., the angle between the neutrino direction and the launch vector. 

The launch vector $\vec{l}$ corresponds to the incoming signal direction after correcting for the bending in the firn. In practice, almost all neutrinos observed with an Askaryan detector will have the interaction vertex below the firn (see example 1 of \cite{NuRadioMC}). Thus, the exact location of the interaction vertex is not needed to correct the incoming signal direction and polarization for the bending in the firn. 

In summary, a measurement of the incoming signal direction, polarization and viewing angle is required to determine the neutrino direction. The incoming signal direction can be reconstructed precisely from the pulse arrival times in multiple channels yielding a resolution of better than \SI{1}{\degree} (see e.g. \cite{COSPAR2018,GaswintICRC2019}). The viewing angle is difficult to measure but is anyway constraint to a few degrees because of the narrowness of the Cherenkov cone (cf. last part of Eq.~(\ref{eq:all})). A detailed MC simulation using NuRadioMC \cite{NuRadioMC} yields a scatter of $\sigma_\Theta = \SI{4}{\degree} (\SI{3}{\degree})$ for neutrino energies of \SI{e18}{eV} (\SI{e17}{eV}). 
The precision of the polarization reconstruction depends strongly on the experimental setup and can range from a few degrees to being largely unconstrained. 

\begin{figure}[t]
    \centering
    \includegraphics[width=0.7\textwidth]{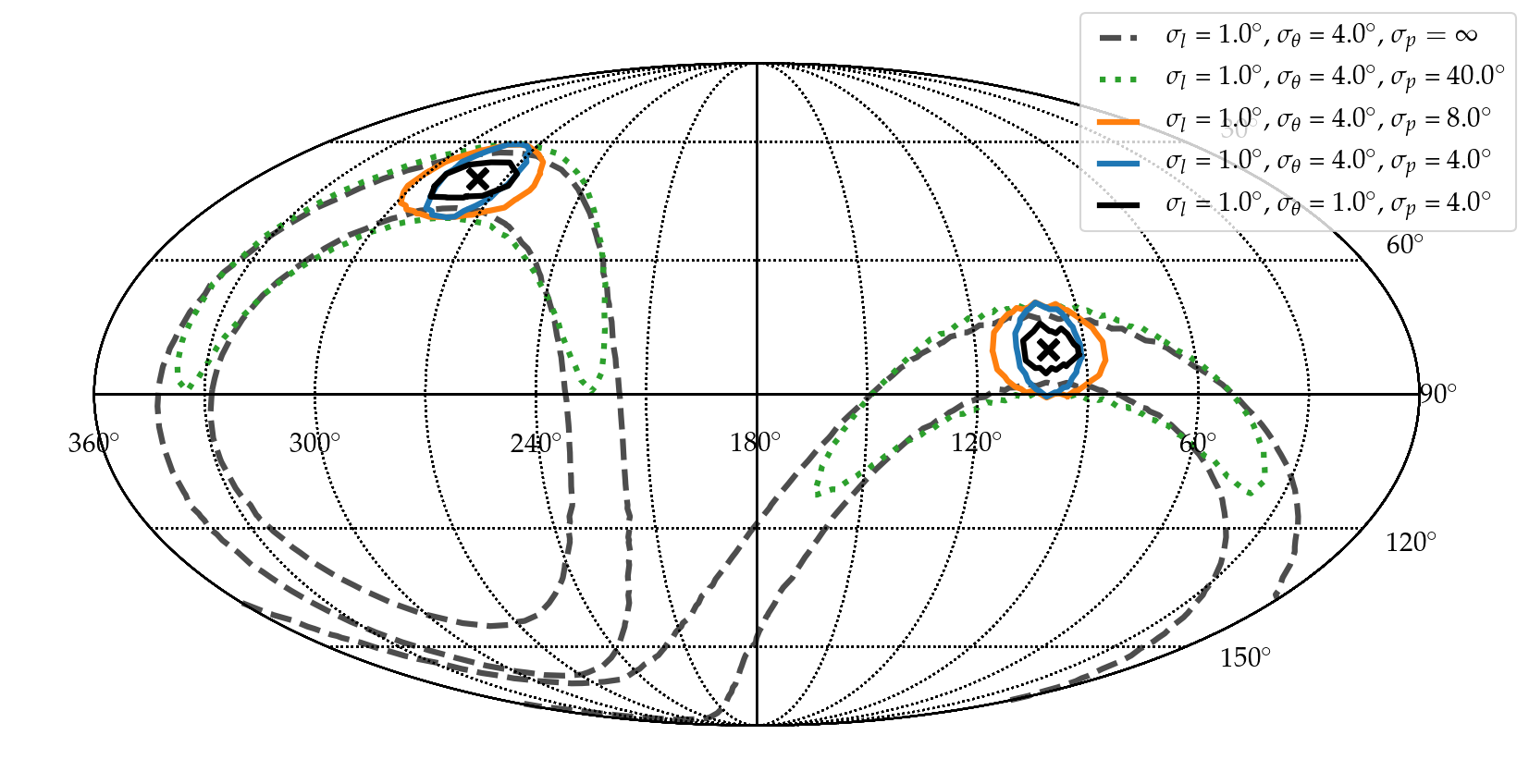}
    \caption{Skymap in local coordinates. The crosses indicate two hypothetical neutrino directions. The lines show the 90\% CL contours of the reconstructed neutrino direction for different assumptions on uncertainties.}
    \label{fig:skymap}
\end{figure}

We show the resolution of the neutrino direction for different assumptions on the uncertainties on signal direction, viewing angle and polarization in Fig.~\ref{fig:skymap}. We quantify the resolution by calculating the solid angle covered by the 90\% CL contour and present the results in Tab.~\ref{tab:direction}. 
The banana-like shape of the contours are a consequence of the constraint of the polarization vector being perpendicular to the signal direction. 

\begin{table}[h]
    \centering
    \begin{tabular}{c c c c c}
    \hline \hline 
        $\sigma_l$ & $\sigma_\theta$ & $\sigma_p$ & $\sigma_{68\%}$ & $A_{90\%\, \mathrm{CL}}$  \\ \hline
        \SI{0.2}{\degree}& \SI{1}{\degree} & \SI{2}{\degree} & \SI{2.1}{\degree} & \SI{0.01}{sr} \\
         \SI{1}{\degree}& \SI{1}{\degree} & \SI{4}{\degree} & \SI{3.9}{\degree} & \SI{0.02}{sr} \\
         \SI{1}{\degree}& \SI{4}{\degree} & \SI{4}{\degree} & \SI{5.7}{\degree} &\SI{0.06}{sr} \\
         \SI{1}{\degree}& \SI{4}{\degree} & \SI{8}{\degree} & \SI{8.3}{\degree} & \SI{0.12}{sr} \\
         \SI{1}{\degree}& \SI{4}{\degree} & \SI{40}{\degree} & - & \SI{0.60}{sr} \\
         \SI{1}{\degree}& \SI{4}{\degree} & unconstrained & - & \SI{1.23}{sr} \\ \hline \hline 
    \end{tabular}
    \caption{Resolution of neutrino direction quantified as solid angle of the 90\% CL contour ($A_{90\%\, \mathrm{CL}}$) as a function of uncertainty of the signal direction $\sigma_l$, viewing angle $\sigma_\theta$ and polarization $\sigma_p$. For small uncertainties where the resulting neutrino direction uncertainty can be approximated with a 2d Gaussian distribution, we also specify the $\sigma$ parameter.}
    \label{tab:direction}
\end{table}

\subsection{Inelasticity}
\label{sec:y}
\begin{figure}[t]
    \centering
    \includegraphics[width=0.49\textwidth]{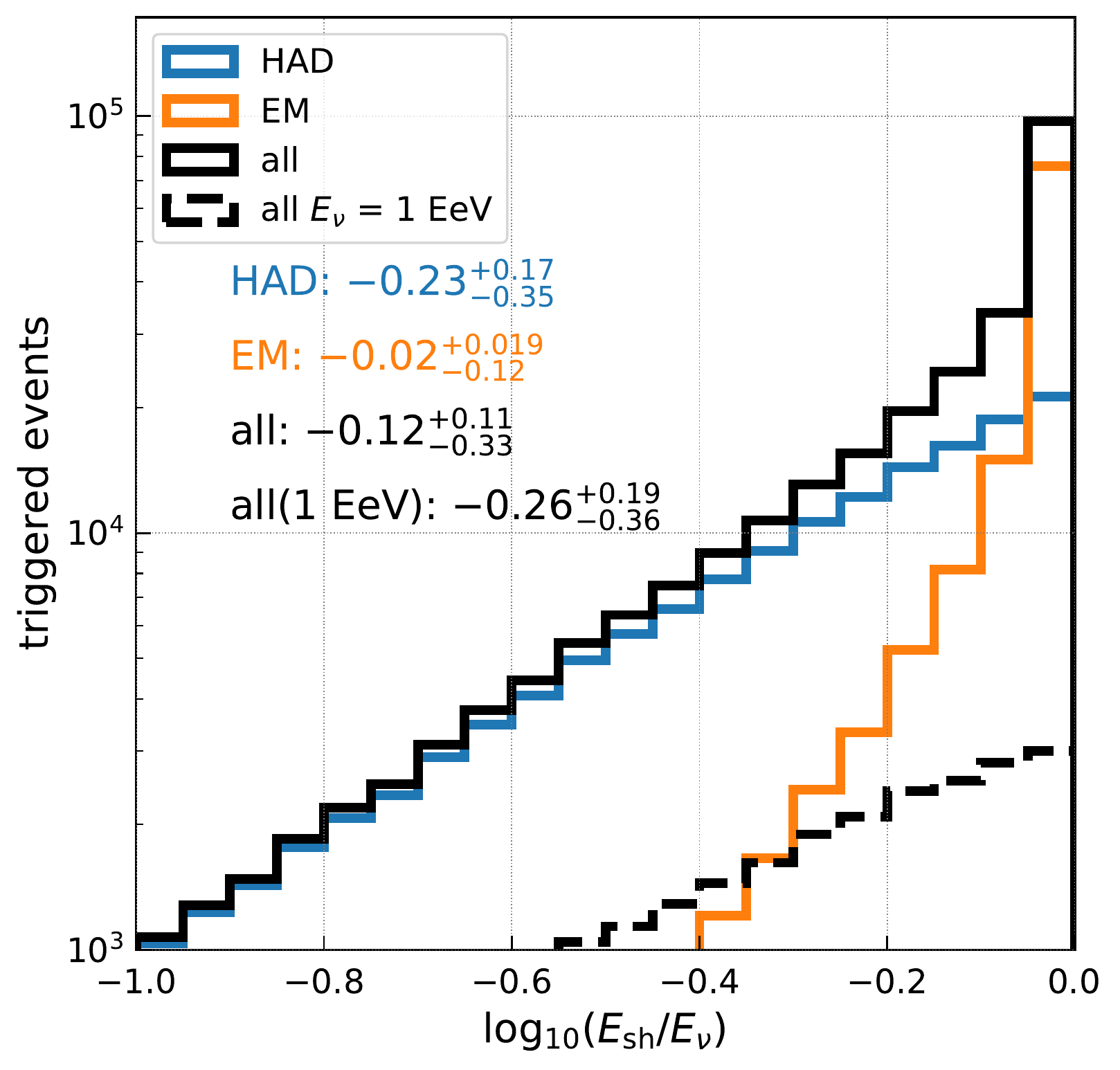}
    \includegraphics[width=0.49\textwidth]{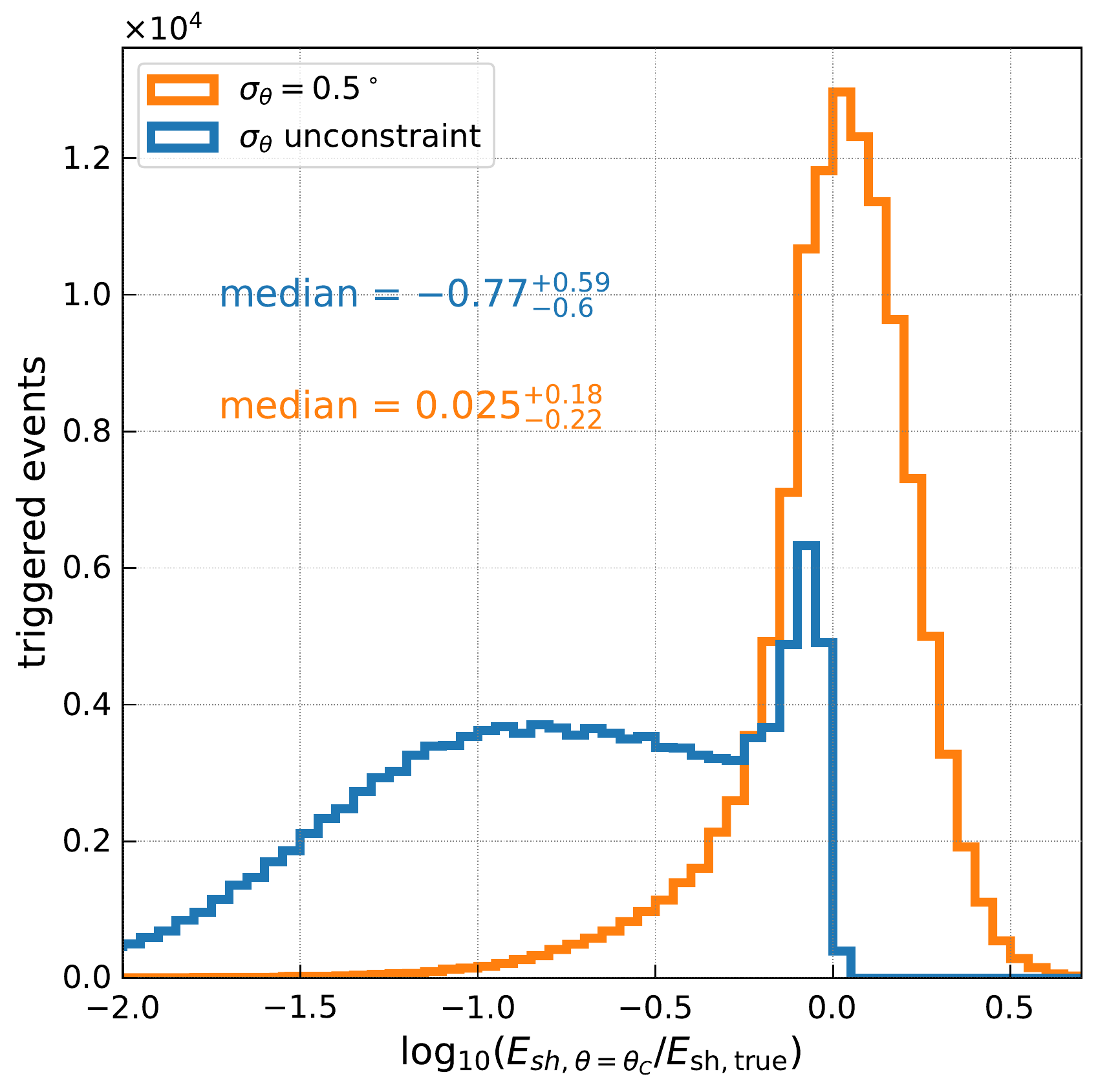}
    \caption{(left) Ratio of shower energy and neutrino energy of triggered events for a $3\, V_\mathrm{RMS}$ trigger at a noise temperature of \SI{300}{K} and a neutrino energy spectrum following a $E^{-2.2}$ power law as measured by IceCube plus a GZK signal expectation. (right) Scatter of reconstructed shower energy if the viewing angle is unknown (blue) and for a \SI{0.5}{\degree} uncertainty (orange).}
    \label{fig:ylimit}
\end{figure}

The second term of Eq.~(\ref{eq:all}) ($|\vec{\varepsilon}_0|$) represents the dependence of the Askaryan signal on the neutrino energy. The Askaryan signal amplitude does not depend directly on the neutrino energy by scales linearly with the shower energy. The fraction of energy transferred into the shower (the inelasticity) is a stochastic process which limits the achievable energy resolution and is presented in Fig.~\ref{fig:ylimit} left which was obtained from a detailed NuRadioMC simulation \cite{NuRadioMC, DNR2019} for a GZK + $E^{-2.2}$ neutrino energy spectrum. The distribution is biased towards events with a high energy transfer into the shower. This uncertainty is energy dependent: The larger the energy the larger is the resulting scatter from inelasticity variations because the data set is less biases to high energy transfers (see curve for \SI{1}{EeV} in Fig.~\ref{fig:ylimit} left). 
As the distribution is strongly asymmetric, we estimate the uncertainty with the 68\% quantiles to about a factor of 2 ($\sim$0.3 in $\log_{10}(E_\mathrm{sh}/E_\nu)$).

This uncertainty from inelasticity sets the scale for the required experimental precision: The uncertainties of other quantities impacting the energy reconstruction should be small enough to not significantly increase the energy uncertainty beyond the inelasticity limit but don't need to be much more precise either.

\subsection{Signal attenuation}
The third term of Eq.~(\ref{eq:all}) describes the attenuation of the radio signal during propagation. The signal amplitude decreases proportional to the distance $R$ to the neutrino vertex. In addition, the signal is attenuated exponentially as $e^{-R/L(f)}$ where $L(f)$ is the frequency dependent attenuation length with typical values ranging from \SI{500}{m} to \SI{3}{km}. 

\subsection{Viewing angle}
The last term of Eq.~(\ref{eq:all}) describes the dependence on the viewing angle $\theta$. If the shower is observed at the Cherenkov angle $\theta_C = \cos^{-1}(1./n_\mathrm{ice}) \approx \SI{55.8}{\degree}$, the emission along the shower track adds up coherently yielding the largest signal amplitude.

To get an upper bound on the influence of the viewing angle on the reconstructed shower energy, we calculate the scatter of the shower energy assuming no information at all on the viewing angle. We use the same NuRadioMC simulation as in Sec.~\ref{sec:y} for a GZK + $E^{-2.2}$ neutrino energy spectrum and correct the signals of all triggered events assuming an electromagnetic shower that was observed on the Cherenkov cone. This results in the distribution of $\log_{10}(E_\mathrm{rec}/E_\mathrm{true})$ shown in Fig.~\ref{fig:ylimit} right with a 68\% quantile of $\sim$0.6 that corresponds to a factor of 4 on a linear scale. 

\section{Experimental determination of quantities}
In summary, the determination of the neutrino energy and direction requires a measurement of the signal amplitude, the vertex distance, the polarization and the viewing angle. This section discusses how these quantities can be measured experimentally. 

\subsection{Measurement of signal arrival direction}
The signal arrival direction is determined from the signal arrival time measured in several spatially separated antennas. This is a well studied problem in the literature and the achievable resolution depends on the lever arm (the separation between antennas) and the time resolution. With the ARIANNA detector, a precision of less than a degree was achieved for an in-situ calibration measurement where pulses where emitted from a transmitter deep in the ice \cite{COSPAR2018, GaswintICRC2019}. We note that this measurement also showed that the bending of signal trajectories in the firn can be corrected for accurately. 

\subsection{Measurement of signal polarization}
To determine the signal polarization, the Askaryan radio pulse needs to be measured in several antennas with complementary polarization response. 
The optimal case is to have (multiple) antennas with equal gain to orthogonal polarization components and to use the same antenna type to minimize systematic uncertainties. This can be achieved by deploying broadband high-gain LPDA antennas with different orientations (as proposed for future large-scale Askaryan detectors e.g. \cite{COSPAR2018}). 

We tested the polarization reconstruction via the measurement of the more abundant cosmic-ray air showers that produce a radio signal very similar to the neutrino signal. First, we studied the polarization reconstruction in a thorough Monte Carlo study. We developed a new \emph{forward folding} technique where we optimize the parameters of an analytic model of the electric-field pulse by minimizing the difference between the measured voltages at the antennas and the predicted pulses \cite{NuRadioReco}. This leads to a significant improvement in precision compared to the standard way of unfolding the antenna response simultaneously from the measured voltages. 
We simulated the dedicated cosmic-ray station of the ARIANNA detector consisting of four upward facing LPDA antennas where two antennas are oriented along Easting and the other pair along Northing. For a realistic distribution of signal-to-noise ratios and with a reasonable assumption on achievable systematic experimental uncertainties, we obtain a polarization resolution of \SI{2.6}{\degree} \cite{NuRadioReco, NellesICRC2019}. 

Then, we analyzed data from the ARIANNA cosmic-ray station and find scatter between reconstructed and theoretically predicted polarization of \SI{7}{\degree} (see \cite{NellesICRC2019} for more details). In the future, we expect to reduce the polarization resolution significantly  with a better calibration (e.g. the antenna positions and amplifier responses) to about \SI{2}{\degree}-\SI{3}{\degree}.

For neutrinos, one can envision a better polarization reconstruction because the Askaryan signal is measured in more antennas resulting in less distortion due to noise.

\subsection{Measurement of distance to neutrino interaction vertex}

The distance from the observer to the neutrino interaction vertex can be measured experimentally using the D'n'R technique \cite{COSPAR2018, DNR2019,Allison2019}: An antenna placed $\sim\SI{15}{m}$ below the surface will observe two Askaryan pulses for most neutrino events \cite{NuRadioMC}, one signal from a direct path to the antenna, and a second delayed signal that is reflected off the ice surface. For most geometries we get total-internal-reflection at the ice-air boundary leading to two pulses with equal amplitude. At deeper depths though the efficiency to detect both pulses reduces quickly \cite{NuRadioMC}. Therefore, this technique can only be exploited efficiently with a shallow detector. The vertex distance is a function of the time delay between the two pulses $\Delta t$ and the incoming signal direction.

The main advantage of the D'n'R technique -- compared to a 3D detector that measures the curvature of the wavefront from the signal times in multiple spatially separated antennas -- is that the time delay $\Delta t$ can be measured very precisely. An in-situ measurement with the ARIANNA detector showed a time resolution of \SI{80}{ps} \cite{DNR2019}. 

\begin{figure}[t]
    \centering
    \includegraphics[width=0.47\textwidth]{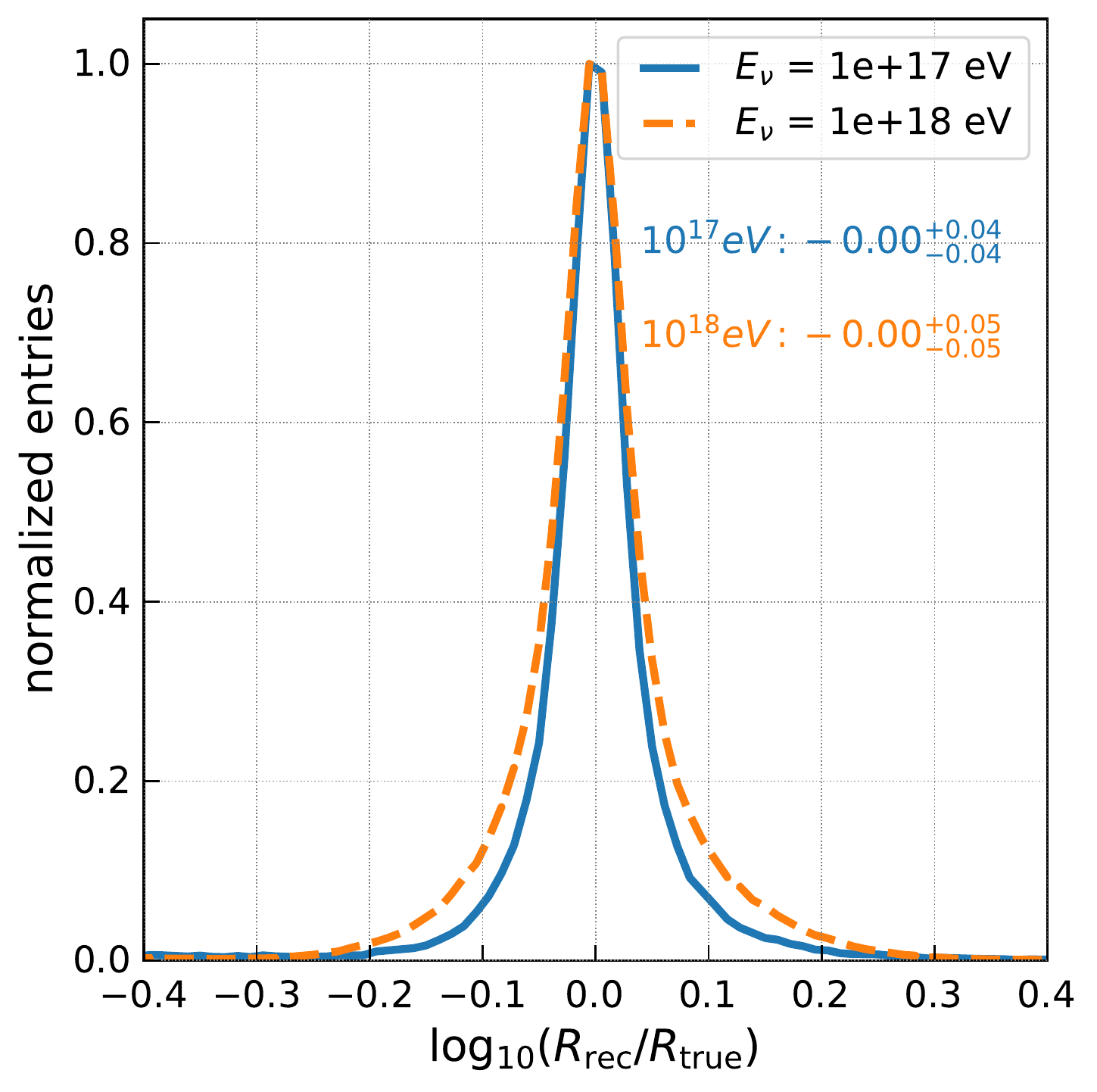}
    \includegraphics[width=0.47\textwidth]{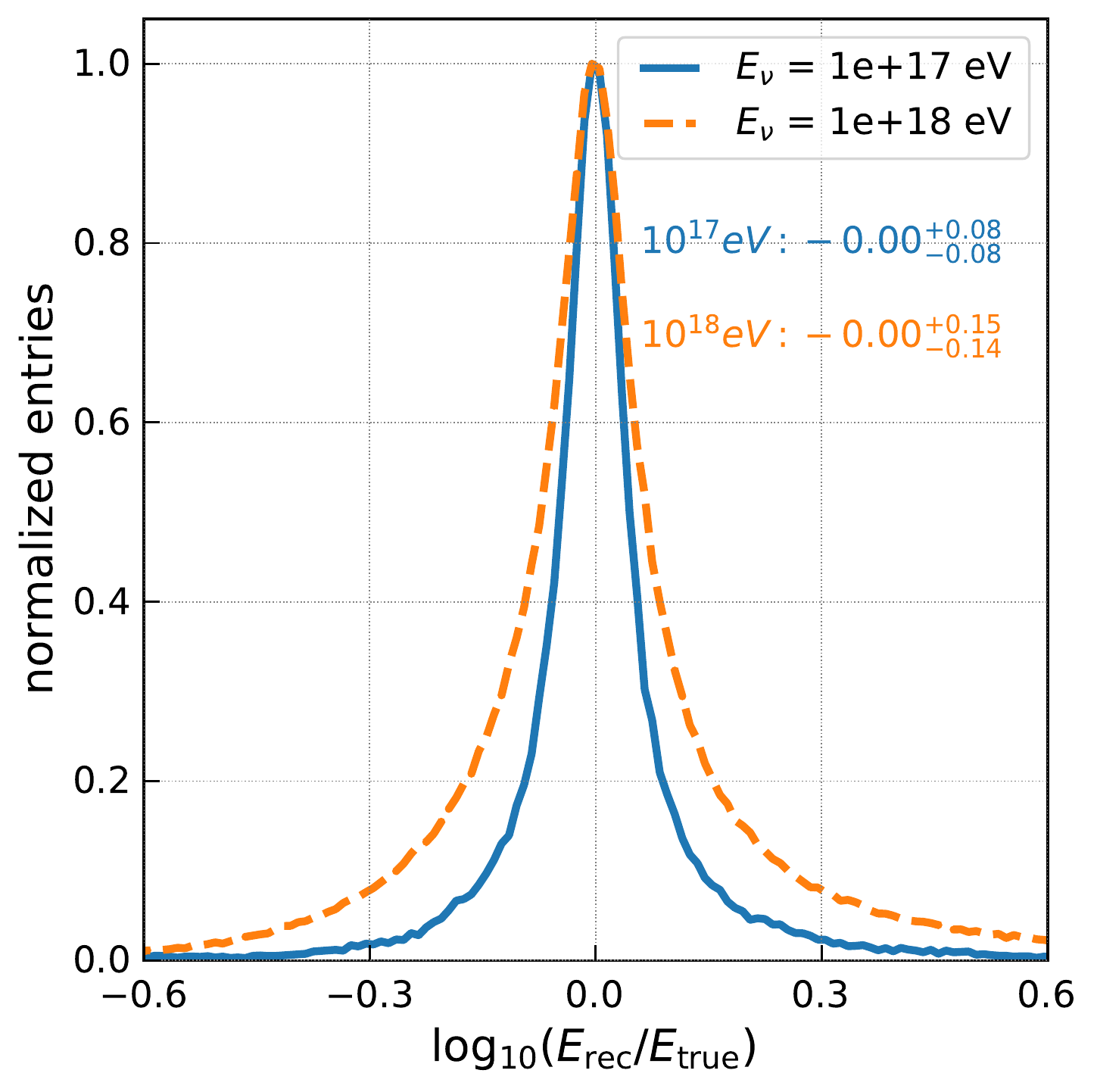}
    \caption{(left) Vertex distance resolution for a \SI{-15}{m} deep receiver and uncertainties of \SI{0.2}{ns} in the D'n'R time delay and \SI{0.2}{\degree} in the zenith direction. (right) Contribution to the energy resolution from uncertainties of the vertex distance. From \cite{DNR2019}.}
    \label{fig:vertex}
\end{figure}

We used NuRadioMC to simulate the achievable vertex resolution and corresponding contribution to the energy uncertainty. For a uncertainty in $\Delta t$ of \SI{0.2}{ns} and \SI{0.2}{\degree} in the zenith angle of the incoming signal direction (as demonstrated by ARIANNA \cite{GaswintICRC2019}) we obtain the vertex and energy resolution presented in Fig.~\ref{fig:vertex}. The resolution depends on the energy itself. At higher neutrino energies the vertices are typically further away which leads to larger uncertainties. At $E_\nu = \SI{e17}{eV} (\SI{e18}{eV})$ we find a vertex distance resolution of 10\% (12\%). This translates into a contribution to the energy uncertainty of 20\% (40\%) which is well below the limit from unknown inelasticity. Reaching a similar resolution in the vertex distance reconstruction will be challenging for a deep 3D detector.

\subsection{Measurement of viewing angle}
The viewing angle can be measured via two complementary techniques: First, via a mapping of the Cherenkov cone via the measurement of the Askaryan signal in multiple antennas that observe the shower under different viewing angles. This requires a sufficient spatial separation between the antennas. The optimal spacing will depend on the vertex distance, hence, it will be difficult to optimize a detector layout equally well for all possible neutrino events.

Second, the viewing angle can be determined by measuring the frequency spectrum of the Askaryan signal. The frequency spectrum of Askaryan pulses increases in amplitude with frequency up a cutoff frequency. The cutoff frequency depends on the viewing angle. It is highest ($> \SI{1}{GHz}$) at the Cherenkov angle and decreases with increasing deviation from the Cherenkov angle. Thus, the requirement for the detector to measure the viewing angle is a broadband frequency response which is provided by the LPDA antennas of the shallow station design. 

We again use cosmic rays to estimate the energy uncertainty due to the viewing angle. For cosmic rays, the energy depends on first order only on the viewing angle, incoming signal direction and polarization. This is because the atmosphere is transparent to radio waves and because for a fixed zenith angle, the air shower has a fixed distance to the observer (neglecting $X_\mathrm{max}$ fluctuations). Hence, the energy reconstruction of cosmic rays with a single radio detector station is a test of the influence of the viewing angle reconstruction in the neutrino energy reconstruction. 
In \cite{Welling2019}, a method was presented to reconstruct the cosmic-ray energy from the radio signal measured in one single station that built up on the novel forward folding technique \cite{NuRadioReco} to recover the incident electric-field pulse (i.e. the polarization and frequency spectrum) from the measurement of individual antennas. The shape of the frequency spectrum then correlates with the viewing angle. In a detailed simulation study, a cosmic-ray energy resolution of 15\% was obtained on a realistic Monte Carlo set. 

The same forward folding method can be applied to neutrinos after modifying electric-field description to match the Askaryan signal of neutrinos. We expect that a similar resolution on the viewing angle can be achieved for neutrinos. Thus, the contribution from viewing angle uncertainties on the neutrino energy resolution will be much smaller than the contribution from inelasticity and vertex distance. 

\section{Conclusions}
We have explored how the neutrino direction and energy can be determined from an Askaryan detector. We discussed which low level quantities need to be measured and evaluated their impact on the direction and energy resolution. This serves as general guidelines for optimizing a detector layout. We also estimated the uncertainties for a shallow detector design. 

To determine the neutrino direction, the signal arrival direction and polarization as well as the viewing angle need to be measured. With a well calibrated shallow detector with multiple LPDA antennas of different orientation, a resolution of about \SI{2}{\degree} is achievable where the polarization measurement was identified as the limiting factor. 

To determine the neutrino energy, a measurement of the signal strength, the polarization, the vertex distance and the viewing angle is required. A shallow detector design is capable of measuring all these properties to better than the physical limit imposed by inelasticity variations. The fluctuations of how much energy is transferred into the particle shower limits the energy resolution to about a factor of 2. 
The vertex distance can be measured precisely via the D'n'R technique, and the viewing angle can be determined from the frequency spectrum of the Askaryan pulse. 

\section{Acknowledgements}
We are grateful to the U.S. National Science Foundation-Office of Polar Programs, the U.S.
National Science Foundation-Physics Division (grant NSF-1607719) and the U.S. Department of Energy. We thank generous support from the German Research Foundation (DFG), grant NE 2031/2-1 and GL 914/1-1, the Taiwan Ministry of Science and Technology. D. Besson and A. Novikov acknowledge support from the MEPhI Academic Excellence Project (Contract No. 02.a03.21.0005) and the Megagrant 2013 program of Russia, via agreement 14.12.31.0006 from 24.06.2013 
\bibliographystyle{JHEP}
\bibliography{bib}

\end{document}